\documentstyle[12pt]{article}
\def\al{\alpha}
\def\be{\beta}
\def\ga{\gamma}
\def\de{\delta}

\def\et{\eta}
\def\th{\theta}

\def\si{\sigma}

\def\ph{\phi}

\def\ch{\chi}
\def\ps{\psi}

\def\Ga{\Gamma}
\def\De{\Delta}

\def\fr#1#2{{{#1} \over {#2}}}
\def\prt{\partial}

\def\vev#1{\langle {#1}\rangle}

\def\bracket#1#2{\langle{#1}|{#2}\rangle}

\def\half{{\textstyle{1\over 2}}}
\def\frac#1#2{{\textstyle{{#1}\over {#2}}}}

\def\lsim{\mathrel{\rlap{\lower4pt\hbox{\hskip1pt$\sim$}}
    \raise1pt\hbox{$<$}}}
\def\gsim{\mathrel{\rlap{\lower4pt\hbox{\hskip1pt$\sim$}}
    \raise1pt\hbox{$>$}}}
\def\sqr#1#2{{\vcenter{\vbox{\hrule height.#2pt
         \hbox{\vrule width.#2pt height#1pt \kern#1pt
         \vrule width.#2pt}
         \hrule height.#2pt}}}}

\def\oc{\widehat{\cos\ph}}
\def\os{\widehat{\sin\ph}}

\newcommand{\beq}{\begin{equation}}
\newcommand{\eeq}{\end{equation}}
\newcommand{\bea}{\begin{eqnarray}}
\newcommand{\eea}{\end{eqnarray}}
\newcommand{\rf}[1]{(\ref{#1})}

\renewenvironment{thebibliography}[1]
 { \rm
   \begin{list}{\arabic{enumi}.}
    {\usecounter{enumi} \setlength{\parsep}{0pt}
     \setlength{\itemsep}{3pt} \settowidth{\labelwidth}{#1.}
     \sloppy
    }}{\end{list}}

\begin{document}
\titlepage

\begin{flushright}
{IUHET 295\\}
{COLBY-95-01\\}
\end{flushright}
\vglue 1cm

\begin{center}
{{\bf ELLIPTICAL SQUEEZED STATES AND RYDBERG WAVE PACKETS \\}
\vglue 1cm
{Robert Bluhm,$^a$ V. Alan Kosteleck\'y,$^b$ and
Bogdan Tudose$^b$\\}
\bigskip
{\it $^a$Physics Department\\}
\medskip
{\it Colby College\\}
\medskip
{\it Waterville, ME 04901, U.S.A.\\}
\bigskip
{\it $^b$Physics Department\\}
\medskip
{\it Indiana University\\}
\medskip
{\it Bloomington, IN 47405, U.S.A.\\}

}
\vglue 0.8cm

\end{center}

{\rightskip=3pc\leftskip=3pc\noindent
We present a theoretical construction
for closest-to-classical wave packets
localized in both angular and radial coordinates
and moving on a keplerian orbit.
The method produces a family of elliptical squeezed states
for the planar Coulomb problem
that minimize appropriate uncertainty relations
in radial and angular coordinates.
The time evolution of these states is studied for orbits
with different semimajor axes and eccentricities.
The elliptical squeezed states may be useful for a
description of the motion of Rydberg wave packets
excited by short-pulsed lasers in the presence of external fields,
which experiments are attempting to produce.
We outline an extension of the method to include certain
effects of quantum defects appearing in the alkali-metal atoms
used in experiments.

}

\vskip 1truein
\centerline{\it To appear in the September 1995
issue of Physical Review A}

\vfill
\newpage

\baselineskip=20pt

{\bf\noindent I. INTRODUCTION}
\vglue 0.4cm

Coherent states
\cite{example}
are candidate quantum-mechanical states for
probing the interface between classical and
quantum mechanics.
Obtaining appropriate coherent states for a
given quantum situation can be a difficult task,
even for apparently simple systems.
For example,
although Schr\"odinger succeeded in
finding exact nonspreading
coherent states for the harmonic oscillator
\cite{schr},
he was unable to find an analogous
solution for the Coulomb potential
\cite{wavemech}.
This problem has been discussed by many authors,
and indeed it is now known for the Coulomb case
that there are no exact solutions representing localized
nonspreading packets following the classical motion
\cite{brown,mostowski,nieto,mcab,gay,nau1,zlatev}.

Given the impossibility of solving the original
Schr\"odinger problem,
one can seek instead
closest-to-classical wave packets for the Coulomb potential
without imposing the condition that the expectation
values follow the exact classical motion for all time.
It is natural to consider first
a simplified version of this problem
in which attention is restricted to the radial part
of the motion,
with a fixed (say, p-state) angular wave function.
Physically,
a wave function of this type results when a short laser pulse
incident on a Rydberg atom
excites a coherent superposition of energy states
with identical angular-momentum quantum numbers.

It turns out that a kind of squeezed state
\cite{kim}
called a radial squeezed state (RSS)
provides a good description
of a closest-to-classical radial packet of this type
\cite{rss,rssqdt}.
Such states minimize the uncertainty relation
in a set of variables for which the radial Coulomb
problem takes a form similar to that of an oscillator.
An oscillator squeezed state
is like a coherent state in that it
follows the classical motion for all time.
However, it does not maintain its shape,
instead having a position-space spread that
oscillates periodically
\cite{nieto2}.
It can be shown that
the corresponding Coulomb RSS has radial motion following
the closest-to-classical motion along a keplerian ellipse.
As is characteristic of a squeezed state,
the RSS exhibits oscillations in the uncertainty
product and ratio as a function of time.

Radial wave packets with p-state angular distributions
have been experimentally produced using short-pulsed
lasers to excite a coherent superposition of Rydberg states
\cite{tenWolde,yeazell1,yeazell2,yeazell3,meacher,wals}.
The resulting localized Rydberg wave packets have
features that match some of those of a classical
particle moving in a Coulomb potential.
For example,
the initial motion is periodic with the classical
keplerian period.
While radial Rydberg wave packets
initially follow the classical motion,
they also exhibit effects due to
quantum-mechanical interference.
For instance,
at times beyond a few classical periods
they collapse and undergo a
cycle of fractional/full revivals
\cite{ps,az,ap,nau2}.
Indeed,
at still larger times, well beyond
the appearance of the first full revival,
the quantum-interference effects cause
radial wave packets to undergo a new cycle of
fractional/full superrevivals
\cite{detunings,sr}.
Eventually,
at times large compared with the superrevival time scale,
the wave packets spontaneously decay or lose their
coherence through other decay processes.

The RSS description reproduces the
primary features in the motion of a Rydberg wave
packet produced by a short-pulsed laser,
including the revival and superrevival structures.
Moreover, the RSS approach has a number of attractive features.
A simple example is the shape of an RSS wave packet,
which,
unlike a simple gaussian,
is asymmetrical along the radial direction.
This asymmetry is desirable
in a theoretical description of a packet
produced with a transform-limited laser pulse exciting a
superposition of unequally spaced Coulomb energy levels.

It is natural to ask whether
this analytical description of the radial Coulomb problem
can be extended to incorporate the angular dependence
characteristic of a classical particle
moving in a keplerian ellipse.
The issue is of particular interest at present
because experiments are currently attempting to produce
Rydberg wave packets that move along elliptical trajectories.

The purpose of the present paper is to
address this question.
We provide here a planar extension
of the RSS construction to a description
of minimum-uncertainty wave packets
moving along a keplerian ellipse,
and we discuss some of the key features
of the solution.
Like the RSS,
the packets derived here are squeezed states.
We call them elliptical squeezed states (ESS).

Unlike purely radial packets,
which have p-state angular distributions,
experimental production of
wave packets localized in angular coordinates
necessarily involves generating
a superposition of $l$ states.
Superpositions of different $l$ states in a manifold
of fixed $n$ have been produced
\cite{yeazell4}.
Although a packet of this kind is indeed localized
in angular variables,
the degeneracy of the $l$ states in hydrogen
means that it is stationary and hence
fails to follow the classical motion.

The production of a wave packet localized
in both radial and angular variables
requires creating a superposition
of both $n$ and $l$ states.
To date,
the only experiments that have detected periodic motion
of a Rydberg wave packet with the classical keplerian period
are those performed with radial packets,
which have fixed $l$.
However,
a technique for producing a superposition of both $n$
and $l$ states has been suggested
\cite{gns}.
The proposal involves making use of
a short electric field to convert
an angular state into a localized Rydberg wave packet.
Once produced,
such a wave packet would move on a circular orbit.
If an additional weak electric field is present,
a wave packet moving on an ellipse of arbitrary
eccentricity could be produced.
Its motion would have features similar to those
of the ESS we present here.

We begin our derivation of the ESS
with a discussion of the planar RSS
appropriate for the keplerian motion
in Sec.\ II.
The next task,
performed in Sec.\ IIIA,
is to obtain appropriate quantum-mechanical
operators describing the angular position and angular
momentum of a particle in a central potential.
In the remainder of Sec.\ III,
we proceed to the construction of
a wave packet,
called a circular squeezed state (CSS),
that minimizes the uncertainty product
for the angular variables,
and we discuss some of its properties.

Section IV provides the construction of the ESS
by minimizing the appropriate uncertainty relations.
At the initial time,
the ESS may be written as the products of RSS with CSS.
Since the ESS naturally involve a
superposition over both $l$ and $n$ quantum numbers,
localization in both the angular and radial coordinates
is achieved.
In this section,
we also obtain expectation values for relevant physical
quantities and discuss the time evolution.
The ESS move with classical keplerian periodicity
along an elliptical orbit of fixed mean eccentricity
and semimajor axis.

Since experiments are often performed in alkali-metal
atoms rather than with hydrogen,
it can be important to allow for modifications due
to quantum defects.
Some of these are treated
in Sec.\ V.
Finally,
Sec.\ VI summarizes our results.

\vglue 0.6cm
{\bf\noindent II. PLANAR RADIAL SQUEEZED STATES}
\vglue 0.4cm

In this section,
we outline the construction of
the radial squeezed states
for motion in a plane.
Although the results differ in detail,
much of the analysis is similar to
ref.\ \cite{rssqdt},
to which the reader is referred for
a more complete treatment.

The RSS construction
begins with the classical effective radial hamiltonian
for the Coulomb potential written
in terms of the radial variables $r$ and $p_r$
and converts it to an oscillator description
in terms of new variables $R$ and $P$.
The resulting classical problem is then quantized,
and wave functions are obtained minimizing the
quantum uncertainty relation obeyed by $R$ and $P$.
These wave functions are the RSS.
One motivation for this construction is that
the minimum uncertainty of the RSS with
respect to oscillator coordinates
incorporates some of the attractive features
of oscillator squeezed states,
while maintaining
the RSS time evolution controlled by the
Schr\"odinger equation with a Coulomb potential.

Classically,
the effective radial planar hamiltonian
for a particle in a Coulomb potential is
\beq
H \equiv
\half p_r^2 + \fr {l^2} {2 r^2} - \fr 1 r
= E
\quad ,
\label{classH}
\eeq
where $p_r$ is the radial momentum and $E$ is
the energy.
For $E<0$,
the radial motion is oscillatory between the values
$r_{1,2}$ given by
$r_{1,2} = a(1 \pm e)$,
where $a=1/2|E|$ is the semimajor axis of the orbital
ellipse and $e = (1-2 l^2 |E|)^{1/2}$ is the eccentricity,
with $l$ the constant classical angular momentum.
The classical orbital period is given by
$T_{\rm cl} = \pi /(2 |E|)^{3/2}$.

The conversion to an oscillator description
is via the variable change
\cite{nieto}
\beq
R = \fr 1 r - \fr 1 {l^2}
\quad ,\qquad
P = p_r
\quad .
\label{bigP}
\eeq
In terms of $R$ and $P$,
the equation $H=E$ becomes
\beq
\half P^2 + \half l^2 R^2 = \half e^2
\quad ,
\label{hameq}
\eeq
which has the form of an energy equation for an oscillator.

Promoting $R$ and $P$ to the status of quantum-mechanical operators
$R = \fr 1 r - \fr 1 {l^2}$ and
$P \equiv p_r = -i(\partial_r + 1/2r)$,
where $[r,p_r] = i$,
we obtain the commutation relation $[R,P] = -i/{r^2}$.
The RSS are the wave functions
$\psi(r)$ minimizing the corresponding uncertainty relation
\beq
\De R \De P \ge \half \vev{\fr 1 {r^2}}
\quad
\label{uncert}
\eeq
at fixed time.

The minimum-uncertainty wave packets are given by
\beq
\psi (r) = N_1 r^{\al} \exp [-\ga_0 r -i \ga_1 r]
\quad ,
\label{2drss}
\eeq
where the normalization constant is
$N_1 = (2 \ga_0 )^{\al + 1}/[\Ga (2 \al + 2)]^{1/2}$.
The parameters $\al$, $\ga_0$, and $\ga_1$
are given in terms of the squeezing
\beq
S = \fr {\De R} {\De P} = \fr {2 (\De R )^2}{\vev{\fr 1 {r^2}}}
\quad
\label{Ssqueeze}
\eeq
and expectations of the operators $1/r$ and $p_r$ by
\beq
\al = \fr 1 S - \half
\quad , \qquad
\ga_0 = \fr 1 S \vev{\fr 1 r}
\quad , \qquad
\ga_1 = - \vev{p_r}
\quad .
\label{parms}
\eeq

The wave functions $\psi(r)$ form
a three-parameter family of planar RSS.
The radial part of the classical motion
is uniquely determined by two parameters
specifying a point in the $r$-$p_r$ phase space.
However,
three parameters are needed to fix
the quantum-mechanical solution $\psi(r)$
because,
in addition to specifying
$\vev{r}$ and $\vev{p_r}$,
it is necessary to provide
information about the spread of the packet.
This can be done,
for instance,
by fixing the mean energy $\vev{H}$.

The RSS discussed in
refs.\ \cite{rss,rssqdt}
can be used to describe a radial Rydberg wave packet
with p-state angular distribution
produced by a single short laser pulse.
The corresponding three RSS parameters
for this situation are determined by matching
the RSS expectation values for $\vev{H}$, $\vev{r}$,
and $\vev{p_r}$ to mean values for the Rydberg wave packet
at the outer apsidal point.
Three experimental quantities control this match.
The frequency of the laser pulse determines the mean
energy of the wave packet.
The time delay after the excitation of the wave packet
determines the location of the packet.
Finally,
the duration of the laser pulse fixes
the spread of the wave packet.

Our purpose in the present work
is to obtain wave packets
(the ESS)
that follow the classical keplerian motion along an ellipse.
This implies incorporating nontrivial angular dependence
into the analysis,
so quantities such as the mean energy depend on both radial
and angular coordinates.
Again,
certain expectation values specify the ESS parameters.
However,
the coordinate dependence is sufficiently intertwined that
it is impossible to treat independently
the determination of the three planar RSS parameters
$\al$, $\ga_0$, and $\ga_1$ of Eq.\ \rf{2drss}.
For this reason,
we defer until Sec.\ IV a discussion of this topic.

\vglue 0.6cm
{\bf\noindent III. CIRCULAR SQUEEZED STATES}
\vglue 0.4cm

The goal of this section is to obtain
the circular squeezed states,
which can be viewed
as squeezed states for angular coordinates in the plane.
Sec.\ IIIA contains some remarks about the
choice of angular coordinates.
Sec.\ IIIB derives the CSS,
while Sec.\ IIIC examines properties of the CSS.

\vglue 0.6cm
{\bf\noindent A. Quantum Angular Variables}
\vglue 0.4cm

The effective angular hamiltonian for planar motion
is the square $L^2$ of the angular momentum
operator $L$.
There are,
however,
subtleties involved in the choice of the corresponding angular
coordinates to use in the quantum description.
This subsection briefly discusses
some of the issues directly relevant to the construction of the CSS
and provides the definitions and conventions we have adopted.
For more details and a guide to the early literature
on this topic,
we refer the reader to the review \cite{cn}.

Among the possible choices for the angular coordinate are
the continuous variable $\ph_c$ with $-\infty < \ph_c < \infty$
and the periodic variable $\ph_p$ with $-\pi \le \ph_p < \pi$.
Both these choices present problems at the quantum level.
The continuous variable $\ph_c$ is not periodic,
and hence is not an observable
in the Hilbert subspace of functions
for which $L= -i \prt_\ph$ is hermitian
under a conventional inner product.
In contrast,
$\ph_p$ is periodic but is discontinous at $\ph_p=\pi$,
which causes unusual effects with derivative operators
such as the angular momentum.
For instance,
direct substitution into the commutation relation $[\ph_p,L]$
generates $\de$-function contributions.

These difficulties can be avoided by choosing instead angular
coordinates that are both periodic and continuous.
However,
a single such quantity is insufficient
to specify uniquely a point on a circle,
since periodicity implies the existence of extrema
and hence no single quantity can be one-to-one.
One possibility
\cite{wl}
is to use two angular coordinates,
$\cos \ph$ and $\sin \ph$,
where $\ph$ is either $\ph_c$ or $\ph_p$.
Classically, this is a natural choice
since it corresponds to the identification
$(x,y) \rightarrow (\cos \ph,\sin \ph)$
on a unit circle.

The corresponding quantum operators can be defined
through their matrix elements in the Hilbert space.
We take as the inner product the definition
\beq
\bracket{\ps_1}{\ps_2}=
\int_{-\pi}^{\pi} d\ph ~\ps_1^* \ps_2
\quad .
\label{innprod}
\eeq
Matrix elements of the quantum operators
$\oc$ and $\os$ are then given by insertion of the
coordinate functions $\cos\ph$ and $\sin\ph$
in the inner product \rf{innprod}.

The angular operators obey the nonlinear relation
$\os^2 + \oc^2 = I$,
where $I$ is the identity operator.
Note that using one of the two angular coordinates
along with the sign of the other
suffices to determine a unique location on a circle.
However,
this choice is not smooth.
Taken together,
the pair of operators $\oc$, $\os$
have all the features needed
for a quantum-mechanical treatment.

The quantum angular-coordinate operators
$\oc$ and $\os$
and the angular momentum $L$
are intertwined by the commutation relations
\beq
[\oc ,L] = -i ~\os
\quad , \qquad
[\os ,L] = i ~\oc
\quad , \qquad
[\oc ,\os ] = 0
\quad .
\label{cr1}
\eeq
The uncertainty relations following from these equations are
\beq
\De \oc \, \De L \ge \half \vert \vev{\os} \vert
\quad ,
\label{ur1}
\eeq
\beq
\De \os \, \De L \ge \half \vert \vev{\oc} \vert
\quad ,
\label{ur2}
\eeq
and
\beq
\De \os \, \De \oc \ge 0
\quad .
\label{ur3}
\eeq
The last of these relations shows
that $\oc$ and $\os$
can be specified simultaneously to arbitrary precision,
which is intuitively reasonable since both
quantities are needed to determine the
location of the quantum particle.

For simplicity,
in the remainder of this paper
we adopt the usual convention
of writing $\oc$ as $\cos\ph$ and $\os$ as $\sin\ph$,
leaving the context to determine whether
an operator or a function is intended.

\vglue 0.6cm
{\bf\noindent B. Construction of CSS}
\vglue 0.4cm

We begin by seeking a minimum-uncertainty state
that is centered about $\ph = 0$.
Circular symmetry then implies the conditions
\beq
\vev{\sin \ph} = 0
\quad , \qquad
\vev{\cos \ph} > 0
\quad .
\label{cond2}
\eeq
This means that the uncertainty relation
\rf{ur2}
provides the only nontrivial constraint.
A wave packet $\chi(\ph)$ minimizing this relation
must obey the differential equation
\beq
(\sin \ph) \chi =
-i \fr 1 \de (L - \vev{L}) \chi
\quad ,
\label{chieq1}
\eeq
where the squeezing $\de$
in the angular coordinates is a real constant given by
\beq
\fr 1 \de \equiv \fr {\De \sin \ph} {\De L}
= \fr {2(\De \sin \ph)^2} {\vert \vev{\cos \ph} \vert}
\ge 0
\quad .
\label{bigS}
\eeq

The solution to Eq.\ \rf{chieq1} is
\beq
\ch(\ph) = N_2 \exp(\de \cos \ph + i \be \ph)
\quad ,
\label{css}
\eeq
where we have defined the real parameter
$\be = \vev{L}$.
Note that $\be$ must be integer for the $\ch(\ph)$
to be single valued.
This is a special case of the general result
that minimum-uncertainty angular packets must have
integer angular-momentum expectations
\cite{kt}.
The normalization constant $N_2$ is given by
\beq
N_2 = \sqrt{\fr 1 {2 \pi I_0 (2 \de)}}
\quad ,
\label{norm2}
\eeq
where $I_n(z)$ is a modified Bessel function
of the first kind.
The two-parameter family of states \rf{css} are the CSS.
They have previously entered the literature
in the context of uncertainty relations
for phase and angle variables
\cite{cn}.
In addition to $\vev{L} = \be$ and $\vev{\sin\th}= 0$,
already imposed above,
we find
\beq
\vev{\cos \ph} = \fr {I_1 (2 \de)} {I_0 (2 \de)} \ge 0
\quad ,
\label{exp2}
\eeq
as expected.

By construction,
the CSS \rf{css} have magnitude peaking at $\ph=0$.
It is shown in Sec.\ IV
that this is appropriate
for use in deriving packets moving along an ellipse
aligned with outer apsidal point at $\ph = 0$,
which is the orientation most likely to be relevant
in future experiments.
However,
modified CSS with magnitude peaking
at some angle $\ph=\ph_0$ instead
would be relevant for
packets moving along an ellipse of arbitrary orientation.
The remainder of this subsection
concerns these modified CSS.

We first remark that imposing $\vev{\cos \ph} < 0$
instead of Eq.\ \rf{cond2}
results in a sign change of the inverse squeezing $\de$
and hence in a packet of the same shape as Eq.\ \rf{css}
but centered about $\ph=\pi$.
Another possibility is
to seek a configuration with
$\vev{\cos \ph} = 0$ instead of \rf{cond2},
whereupon the interesting constraint becomes Eq.\ \rf{ur2}
instead of \rf{ur1}.
The solution is again of the form \rf{css}
but with $\sin \ph$ replaced by $\cos \ph$ in the exponential,
representing a packet centered at
$\ph = \pm \pi/2$
according to the sign of $\vev{\sin \ph}$.

One approach to the construction of a packet centered
at an arbitrary angle $\ph_0$
is to take advantage of rotational covariance.
Implicit in the use of
$\cos \ph$ and $\sin \ph$ as angular variables
is a choice of origin for $\ph$,
or, equivalently,
a choice of orientation for
the coordinate axes in the plane.
However,
the effective angular hamiltonian $L^2$
has U(1)-rotational invariance
corresponding to independence of the choice of origin.
A rotation by an angle $\ph_0$
may be implemented by a
translation in the continuous coordinate,
$\ph_c \rightarrow \ph_c - \ph_0$.
This leaves invariant all physical observables
since the inner product \rf{innprod} is invariant
on functions of the periodic coordinates
$\sin\th$ and $\cos\th$.

The conditions \rf{cond2} and
the solution \rf{css}
break this U(1) symmetry.
A U(1) rotation of \rf{css}
by $\ph \rightarrow \ph - \ph_0$
produces a different packet with maximum centered at $\ph_0$:
\beq
\chi(\ph,\ph_0)
= N_2 \exp\left[\de \cos (\ph-\ph_0) + i \be (\ph-\ph_0)\right]
\quad ,
\label{css2}
\eeq
Instead of Eq.\ \rf{ur1},
this packet minimizes an uncertainty relation
between a new angular-coordinate operator $\cos(\ph-\ph_0)$
and the angular-momentum operator $L$.

An interesting issue is whether there is a meaningful way
to minimize both \rf{ur1} and \rf{ur2}.
Simultaneous extremization of both relations is impossible.
Furthermore,
various combinations could be chosen for minimization,
producing a variety of interpolations between the
solutions found at $\ph = 0, \pm \half\pi, \pi$.
However,
there is a unique prescription for treating both relations
together such that the packet produced is
the rotated version \rf{css2} of \rf{css}.
For any given packet, define the quantities
\bea
\mu^2 &=& [\De\cos(\ph - \ph_0)]^2 (\De L)^2
\nonumber \\
& - &\frac 1 4 |\vev{\sin(\ph - \ph_0)}|^2
\quad , \qquad
\nonumber \\
\nu^2 &=& [\De\sin(\ph - \ph_0)]^2 (\De L)^2
\nonumber \\
& - & \frac 1 4 |\vev{\cos(\ph - \ph_0)}|^2
\quad .
\label{defs}
\eea
Requiring that the quadratic combination $Q=\mu^2 + \nu^2$
be the constant $Q_0$ given by the case $\ph_0=0$
yields the solution \rf{css2}.
The value of $Q_0$ is
\beq
Q_0 = \fr{\de^2} 4
\left(\fr{I_1(2\de)}{I_0(2\de)}\right)^2
\left[
\fr 1 {\de^2}
-\left(\fr{I_1(2\de)}{I_0(2\de)}\right)^2
+\fr{I_2(2\de)}{I_0(2\de)}
\right]
\quad .
\label{Q}
\eeq

\vglue 0.6cm
{\bf\noindent C. Features of CSS}
\vglue 0.4cm

The form \rf{css} of the CSS can be used
to obtain
second-order expectations.
We find
\beq
\vev{\cos^2 \ph}
= \fr {I_1^\prime (2 \de)} {I_0 (2 \de)}
\quad ,
\label{exp5}
\eeq
\beq
\vev{\sin^2 \ph} = \fr 1 {2 \de}
\fr {I_1 (2 \de)} {I_0 (2 \de)}
= \fr 1 {2 \de} \vev{\cos \ph}
\quad ,
\label{exp4}
\eeq
and
\beq
\vev{L^2}
= \fr {\de} 2 \fr {I_1 (2 \de)} {I_0 (2 \de)} + \be^2
= \fr {\de} {2} \vev{\cos \ph} + \be^2
\quad .
\label{exp6}
\eeq
Note that $\vev{\sin^2 \ph} + \vev{\cos^2 \ph} = 1$,
as required.
In the limit $\de \rightarrow 0$,
$\vev{\sin^2 \ph}$ and $\vev{\cos^2 \ph}$ both approach $\half$,
while as $\de \rightarrow \infty$,
$\vev{\sin^2 \ph} \rightarrow 0$ and
$\vev{\cos^2 \ph} \rightarrow 1$.

The uncertainties in the angular coordinates and momentum are
\beq
(\De \cos \ph)^2 =
\fr {I_0 (2 \de) I_1^\prime (2 \de) - \left[I_1 (2 \de)\right]^2}
{\left[I_0 (2 \de)\right]^2}
\quad ,
\label{unccos}
\eeq
\beq
(\De \sin \ph)^2 = \fr 1 {2 \de}
\fr {I_1 (2 \de)} {I_0 (2 \de)}
= \fr 1 {2 \de} \vev{\cos \ph}
\quad ,
\label{uncsin}
\eeq
\beq
(\De L)^2 = \fr {\de} {2}
\fr {I_1 (2 \de)} {I_0 (2 \de)}
= \fr {\de} {2} \vev{\cos \ph}
\quad .
\label{uncL}
\eeq
These confirm that the uncertainty product
\rf{ur2} is minimized.

The uncertainty relations reveal that the CSS
parameter $\de$ represents the angular-momentum spread of
the solution $\chi(\ph)$.
As $\de$ decreases so does $\De L$,
and hence the angular-coordinate width increases.
In the limit $\de \rightarrow 0$,
the normalization constant
$N_2 \rightarrow 1/\sqrt{2 \pi}$,
so $\chi(\ph)$ tends to an angular-momentum eigenstate
$Y_\be (\ph) = (2 \pi)^{-1/2} \exp(i\be\ph)$.
This state has $\De L \rightarrow 0$,
consistent with the limit $\de \rightarrow 0$ of
Eq.\ \rf{uncL}.
Moreover,
$\vev{\sin \ph} \rightarrow 0$ also,
so both sides of the uncertainty relation \rf{ur1}
vanish in this limit.
This avoids the appearance of infinite uncertainty
in $\De \cos \ph$,
which is impossible for a bounded function.
Similarly, as $\de \rightarrow 0$ both sides of
Eq.\ \rf{ur2} vanish.

The interpretation of $\de$ as the angular-momentum spread
is confirmed by an expansion of the packet \rf{css}
in the vicinity of $\ph = 0$:
$\vert \chi(\ph) \vert^2 \propto
\exp[ 2 \de (1 - \half \ph^2 + \cdots )]$.
This shows that to leading order
$\vert \chi(\ph) \vert^2$ has a gaussian dependence
proportional to $\exp (-\de \ph^2)$,
with angular-coordinate standard deviation
$\si = 1/\sqrt{2 \de}$.
It also implies that a CSS of
finite angular-coordinate width satisfies
$|\vev{\cos \ph}| < 1$ along with the conditions \rf{cond2},
which differs from the corresponding classical particle
located on the unit circle at $\cos \ph = 1$, $\sin\ph = 0$.

Given the initial angular location,
a CSS is specified by two quantities
$\de$ and $\be$.
Compared to the corresponding classical problem,
an extra parameter is needed to fix the quantum solution.
Classically,
the motion on a circle is determined by the initial values
of $\sin \ph$, the sign of $\cos \ph$, and $L$.
At the quantum level,
the initial angular position is specified by
$\vev{\sin \ph}$ and the sign of $\vev{\cos \ph}$,
while $\be = \vev{L}$ establishes the angular momentum.
The extra parameter $\de$
controls the angular spread of the packet
and can be fixed through Eq.\ \rf{exp6} by giving $\De L$.
Evidently,
in an experiment attempting to produce a wave packet localized
in the angle $\ph$,
it is insufficient to excite a state of definite $l$.
A single laser pulse alone therefore
cannot excite an angular wave packet.
A superposition of $l$ states with a spread $\De L$
must be produced by turning on additional fields that excite more
than one value of $l$.

Plots of some normalized CSS as a function of $\ph$
are presented in Fig.\ 1.
The configurations all have $\vev{\sin \ph} = 0$,
$\vev{\cos \ph} > 0$,
and $\be = \vev{L} = 30$.
Three different examples are shown,
with $\De L = 0.5$, $1.5$, and $2.5$
corresponding to $\de \simeq 0.804$, $4.757$, and $12.753$,
respectively.
As expected,
increasing $\de$ decreases
the angular-coordinate spread.

\vglue 0.6cm
{\bf\noindent IV. ELLIPTICAL SQUEEZED STATES}
\vglue 0.4cm

This section
discusses the elliptical squeezed states.
Sec.\ IVA constructs them and explains how
a given ESS is specified in terms of physical quantities.
The issue of time evolution is examined in Sec.\ IVB,
where we show that the ESS move along an orbit
close to a classical ellipse.
In Sec.\ IVC,
features of the ESS solutions
and their relationship to other approaches
are considered.

\vglue 0.6cm
{\bf\noindent A. Construction and Specification of ESS}
\vglue 0.4cm

The CSS solution \rf{css}
is a function only of the angular coordinates
and has neither time dependence
nor dependence on the energy quantum number $n$.
One solution of the full planar Coulomb problem
is the the product of a CSS
with a radial energy eigenstate of given $n$.
However,
this produces a stationary state.

The desired closest-to-classical packet
moving on a keplerian orbit can be obtained
by taking advantage of the separability of the full hamiltonian.
Note that the uncertainty relation \rf{uncert} is independent of $l$
because the uncertainty $\De R$ involves the combination
$R - \vev{R}$.
This implies it is possible to minimize simultaneously
the uncertainty relations \rf{uncert} and \rf{ur1}
at a given time.
We can therefore take as an initial state
the product of an RSS $\psi(r)$ and a CSS $\chi(\ph)$,
giving
\bea
\Psi (r,\ph) &=& \psi(r)\chi(\ph)
\nonumber \\
&=& \left(\fr {(2 \ga_0)^{2 \al + 2}}
{2 \pi I_0(2 \de) \Ga(2 \al + 2)}\right)^\half  r^\al
\exp [-(\ga_0 + i \ga_1) r + \de \cos \ph + i \be \ph ]
\quad ,
\label{ess}
\eea
where we have substituted from Eqs.\ \rf{2drss} and \rf{css}.
This is a normalized five-parameter family of ESS.
They represent minimum-uncertainty packets localized
in both radial and angular coordinates
and evolving in time.
As discussed in Sec.\ III,
the choice of the initial angular-coordinate location
is implicit in the CSS construction
and is taken to be $\vev{\sin \ph} = 0$ and $\vev{\cos \ph} > 0$.

Expectation values of operators in the ESS can be calculated
analytically using Eq.\ \rf{ess}.
Some physically useful expectations are as follows:
\beq
\vev{r} = \fr {\al + 1} {\ga_0}
\quad , \quad\quad\quad
\vev{r^2} = \fr {(\al + 1)(2 \al + 3)} {2 \ga_0^2}
\quad ,
\label{rexp}
\eeq
\beq
\vev{p_r} = - \ga_1
\quad , \quad\quad\quad
\vev{p_r^2} = \fr {\ga_0^2} {2 \al} + \ga_1^2
\quad ,
\label{pexp}
\eeq
\beq
\vev{\sin \ph} = 0
\quad , \quad\quad\quad
\vev{\cos \ph} = \fr {I_1(2 \de)} {I_0(2 \de)} > 0
\quad ,
\label{angexp}
\eeq
\beq
\vev{L} = \be
\quad , \quad\quad\quad
\vev{L^2} = \fr {\de} {2} \fr {I_1(2 \de)} {I_0(2 \de)} + \be^2
\quad ,
\label{Lexp}
\eeq
\beq
\vev{H} = \fr {\ga_0 (\ga_0 - 4)} {2 (2 \al + 1)}
 + \fr {\ga_1^2} 2
 + \fr {\ga_0^2} {\al (2 \al +1)}
 \left( \fr {\de} 2 \fr {I_1(2 \de)} {I_0(2 \de)} + \be^2 \right)
\quad .
\label{Hexp}
\eeq
The expectation value for the energy $\vev{H}$ is obtained
using the planar Coulomb hamiltonian
and depends on all five of the parameters
associated with the component RSS and CSS.

The associated uncertainty products are given by
\bea
\De r \De p_r &=& \half \sqrt{\fr {\al + 1} {\al}}
\\
\De \sin\th \De L &=& \half \fr{I_1(2\de)}{I_0(2\de)}
\\
\De \cos\th \De L &=& \half
\fr {I_0 (2 \de) I_1(2\de) I_1^\prime (2 \de)
- \left[I_1 (2 \de)\right]^3}
{\left[I_0 (2 \de)\right]^3}
\quad .
\label{delrdelp}
\eea
The ESS is not a minimum-uncertainty state in $r$ and $p_r$,
which is as expected since it is constructed to minimize
the uncertainty relation \rf{uncert} instead.
For large values of $\al$, however,
$\De r \De p_r \rightarrow \half$.

We consider an initial configuration where the wave packet
is located at the outer apsidal point of an elliptical orbit.
This choice is consistent with the experimental
situation,
where the uncertainty product for radial Rydberg packets
reaches a minimum near the outer apsidal point
\cite{ps}.
The constraints imposed on
$\vev{\sin \ph}$ and $\vev{\cos \ph}$
ensure that the ellipse has semimajor axis aligned along the
$x$ axis of the coordinate system.
The parameters $\al$, $\be$, $\ga_0$, $\ga_1$ and $\de$
can be fixed in terms of
the spread $\De L$ in the angular momentum
and expectations of the radial coordinates
$\vev{r}$ and $\vev{p_r}$,
of the angular momentum $\vev{L}$,
and of the energy $\vev{H}$.

Classically,
$\vev{r}$, $\vev{p_r}$, and $\vev{L}$ determine
the mean location and initial velocity of the packet.
At the quantum level,
two additional physical conditions must be given to determine
completely the ESS.
These conditions represent the initial width of the packet
in the radial and angular coordinates.
For the first,
we set the energy expectation equal to the
mean energy of a Rydberg packet consisting of a
superposition of $n$ states centered on the value $\bar n$.
A packet of this type can be produced by excitation
with a short laser pulse tuned to the mean energy
$E_{\bar n}\equiv -1/2(\bar n - 1/2)^2$.
If the excitation occurs in the presence of external fields,
a superposition of $l$ states can be achieved as well.
The second extra condition is the spread
$\De L$ in this superposition.
The precise field arrangement will determine $\De L$,
although in practice it may be difficult
to determine \it a priori \rm for a given experiment.

Denote by $\bar l$
the average value of $l$ for the superposition
of $l$ states.
Then,
we choose to set the expectation
of the coordinate $r$ equal to
a radial distance to the outer apsidal point
written in terms of these average quantum numbers,
\beq
r_{\rm out} = ({\bar n} - \half)^2 \left( 1 +
\sqrt{1 - \fr {{\bar l}^2} {({\bar n} - \half)^2}} \right)
\quad .
\label{rout}
\eeq
The expectation of the radial momentum is set to zero.
These choices are not uniquely
enforced by experimental considerations
but do ensure that both RSS and CSS are
readily recovered in suitable limits of the ESS.

The full set of conditions sufficient to fix the five
ESS parameters is therefore as follows:
\bea
\vev{r} = r_{\rm out}
\quad , \qquad
\vev{p_r}  &=& 0
\quad , \qquad
\vev{L} = {\bar l}
\quad ,
\nonumber \\
\vev{H} = E_{\bar n}
\quad , \qquad
&&
\sqrt{\vev{L^2} - \vev{L}^2} = \De L
\quad .
\label{fix6}
\eea
These determine $\al$, $\be$, $\ga_0$, $\ga_1$, and $\de$
in terms of the three quantities $\bar n$, $\bar l$, and $\De L$
whose values depend on a particular excitation scheme
using a short-pulsed laser in the presence of external fields.

\vglue 0.6cm
{\bf\noindent B. Evolution of ESS}
\vglue 0.4cm

With parameters matching a Rydberg wave packet
at the outer apsidal point on an elliptical orbit,
the ESS \rf{ess}
may be taken as a minimum-uncertainty initial solution of the
time-dependent Schr\"odinger equation.
Since $\vev{p_r}=0$ and $\vev{L} = \be$ by construction,
for $\be > 0$ the wave packet will begin to move in
the direction of increasing $\ph$.
The geometry of the ensuing orbit
depends on the values of $\bar n$, $\bar l$, and $\De L$.
In the limit ${\bar l} \rightarrow n-1$,
the orbits for the ESS should become more circular in shape,
and the wave packet should propagate at a fixed mean radial
distance from the nucleus.
As ${\bar l} \rightarrow 1$,
the orbit should become highly elliptical,
and the inner apsidal point should approach the origin.
In this case,
radial oscillations should
occur as the particle passes close to the nucleus,
as is observed for $l=1$ RSS
\cite{rss}.

The time evolution of the ESS may be studied
by expanding $\Psi(r,\ph,t)$
in eigenstates $R_{nl}(r)$ of the energy
$E_n = - 1/2(n-1/2)^2$
and $Y_l(\ph)$ of the angular momentum,
given by
\beq
R_{nl}(r) = N r^{|l|} e^{- \fr r {n-1/2}}
L_{n - |l| -1}^{2 |l|} \left(\fr {2r} {n-\half}\right)
\quad , \qquad
Y_l(\ph) = \fr 1 {\sqrt{2\pi}} \exp(-il\ph)
\quad ,
\label{radwave}
\eeq
where $N$ is a normalization constant,
$n$ is a positive integer,
and $l$ is a positive or negative integer
satisfying $|l| \le (n-1)$.
The expansion of the ESS is
\beq
\Psi(r,\ph,t) = \sum_{n,l} c_{nl} R_{nl}(r) Y_l (\ph)
e^{-i E_n t}
\quad .
\label{expansion}
\eeq
Note that the time dependence of the phase means that
the ESS is separable only at $t=0$.

The expansion coefficients
$c_{nl} = \vev{\Psi(r,\ph,0) \vert R_{nl}(r) Y_l (\ph)}$
may be calculated using Eq.\ \rf{ess} as the initial
wave function $\Psi(r,\ph,0)$.
The result for $l\ge 0$ is
$$c_{nl} = \left[ \fr {(2 \ga_0)^{2 \al + 2}}
 {I_0(2 \de) \Ga(2 \al + 2)} \right]^{\half}
\left[ \fr 1 {(2n-1)} \left( \fr 2 {n-\half} \right)^{2l+2}
\fr {\Ga (n-l)} {\Ga (n+l)} \right]^{\half}
I_{\be -l} (\de)
$$
\beq
\quad
\times
\sum_{p=0}^{n-l-1} \left( \fr {-2} {n-\half} \right)^p
\fr {\Ga(n+l) \Ga(\al+l+p+2)} {p! \Ga(p+2l+1) \Ga(n-l-p)}
\left(\fr 1 {n - \half} + \ga_0 + i \ga_1 \right)^{-(\al+l+p+2)}
\quad .
\label{mess}
\eeq

As an example,
consider a wave packet with ${\bar n}=45$, ${\bar l}=30$,
and $\De L=2.5$.
Using these values and Eq.\ \rf{fix6},
we obtain the ESS-parameter values
$\al \simeq 57.408$, $\be=30$, $\ga_0 \simeq 0.01697$,
$\ga_1 = 0$, and $\de \simeq 12.753$.
The series in \rf{expansion} may be well approximated by
truncating the sum to a finite number of terms centered
on $\bar n$ and $\bar l$.
This permits the ESS to be plotted as a function of $r$
and $\ph$ at different times $t$.

Figure 2a shows the initial ESS defined at $t=0$.
It is located on the positive $x$ axis at the outer apsidal point
and is evidently localized in both radial and angular coordinates.
The radial distance to the outer apsidal point is
$r_{\rm out} \simeq 3443$ in atomic units.

The classical orbital period for motion on a keplerian
ellipse is $T_{\rm cl} = 2 \pi ({\bar n} - 1/2)^3$.
With ${\bar n}=45$,
we obtain $T_{\rm cl} \simeq 13.4$ psec.
Figure 2b shows the ESS at the time
$t = \fr 1 3 T_{\rm cl}$.
It has moved in the direction of positive $\ph$
and is spreading along the elliptical orbit.
Since the packet moves slowly near the outer apsidal point
but faster near the inner apsidal point,
in accordance with Kepler's laws,
it has traversed less than one third of the orbit
at $t=\fr 1 3 T_{\rm cl}$.

Figure 2c shows the ESS at $t= \fr 1 2 T_{\rm cl}$.
It is moving rapidly and has substantial spread along
the elliptical orbit.
Since ${\bar l}=30$ is relatively large for a state with
${\bar n}=45$,
the ESS remains at some distance from the nucleus
and therefore exhibits none of the radial oscillations
seen for RSS p states
\cite{rss}.
The radial distance to the inner apsidal point is
$r_{\rm in} \simeq 517$ a.u.

Figures 2d and 2e show the wave packet at the times
$t = \fr 2 3 T_{\rm cl}$ and $t=T_{\rm cl}$,
respectively.
The motion is slower again as the outer
apsidal point is approached,
and the packet becomes more localized again.
At $t=T_{\rm cl}$,
the wave packet closely resembles the initial wave packet.
However,
the motion is not exactly periodic.
For times $t \gg T_{\rm cl}$,
the wave packet collapses and a cycle of revivals
and superrevivals commences.

The localization in the angle $\ph$ oscillates
as the wave packet goes through a classical orbital cycle.
At $t=0$,
it is highly localized in angle,
while at $t = \half T_{\rm cl}$,
the wave packet is extended around much of the orbit.
By $t = T_{\rm cl}$,
the wave packet has localized again.
This oscillation is characteristic of a squeezed state.
The radial uncertainties are oscillating also,
although not as noticeably as for a p-state radial
wave function.

In Fig.\ 3,
two ESS with different values
of $\bar l$ and hence different orbital eccentricities
are compared.
Both packets have ${\bar n}=45$ and $\De L = 2.5$.
They are viewed at $t = \half T_{\rm cl}$ from a point
on the positive $x$ axis looking towards the nucleus.
Figure 3a,
with ${\bar l}=30$,
is the same as Fig.\ 2c
but viewed from a different perspective.
The ellipticity of the orbit is manifest.
Figure 3b shows a different ESS with ${\bar l}=40$,
for which
$\al \simeq 20.412$,
$\be = 40$, $\ga_0 \simeq 0.00752$, $\ga_1 = 0$,
and $\de \simeq 12.753$.
Since $\De L$ is unchanged,
$\de$ remains the same.
As expected,
the orbit is closer to circular.
For this case,
$r_{\rm in} \simeq 1113$ a.u.,
which is more than twice
the value for the ${\bar l}=30$ wave packet in Fig.\ 3a.

\vglue 0.6cm
{\bf\noindent C. Features of ESS}
\vglue 0.4cm

The literature contains examples
of quantum wave packets other than the ESS
that are nonetheless localized on
classical elliptical orbits.
In this subsection,
we discuss a few additional features of the ESS
and provide a comparison with some of
these other approaches
\cite{gay,nau1}.

The methods of refs.\ \cite{gay,nau1}
both involve the Runge-Lenz vector.
In the classical Coulomb problem,
this vector is a conserved quantity
with magnitude equal to the eccentricity
of the orbit and with
orientation along the semimajor axis pointing from
the focus to the inner apsidal point.
At the quantum level,
the Runge-Lenz operator
in atomic units takes the form
\beq
{\vec A} = \half ({\vec p} \times {\vec L} -
{\vec L} \times {\vec p}) - \fr {\vec r} r
\quad .
\label{qRL}
\eeq

In ref.\ \cite{gay},
the Runge-Lenz operator is scaled by a factor involving
the energy operator at a fixed value of $n$.
Within a fixed $n$ manifold,
the Coulomb problem exhibits an SO(4) symmetry.
Coherent states for the SO(4) symmetry can be
constructed using the scaled Runge-Lenz operator and
the angular-momentum operator.
The resulting wave packets are localized on
elliptical orbits.
However,
they are stationary states that do not follow the
classical motion.

A similar approach is taken
in ref.\ \cite{nau1}.
Planar motion is considered,
and a subset of the operators is used to form an
o(3) algebra.
The operators $A_x$, $A_y$, and $L$ satisfy
\beq
[A_x,A_y] = -2iHL
\quad ,
\label{RLalg}
\eeq
where $H$ is the hamiltonian.
The associated uncertainty relation is
\beq
\De A_x \, \De A_y \ge \vert \vev{HL} \vert
\quad .
\label{RLunc}
\eeq
Quantum solutions minimizing this relation are found.
The resulting states are also stationary
and are elliptical in shape.
By taking a gaussian-weighted superposition of these
states with different values of $n$,
a localized wave packet is produced that moves along a
keplerian orbit and follows the classical motion.
In the limit of large angular quantum numbers,
Eq.\ \rf{RLunc} approximately holds
for these superpositions.

The ESS are different from the solutions of
refs.\ \cite{gay,nau1}.
They naturally contain
a superposition of $n$ and $l$ states,
and they are minimum-uncertainty solutions for both
radial and angular operators.
The shape of the elliptical orbit is
determined by the parameters of the ESS,
which in turn depend on
the experimentally determined
parameters $\bar n$, $\bar l$, and $\De L$.

A natural question to ask is whether the ESS
also provide a minimum-uncertainty solution
of Eq.\ \rf{RLunc}.
For the example plotted in Fig.\ 2,
the values of the ESS parameters given in Sec.\ IV
can be used to determine
numerically the left and right hand sides of \rf{RLunc}.
We find $\De A_x \, \De A_y \simeq 0.1214$,
while $\vert \vev{HL} \vert \simeq 0.0099$ in atomic units.
Defining the quantity
\beq
Z = \fr {\De A_x \, \De A_y - \vert \vev{HL} \vert}
{\vert \vev{HL} \vert}
\quad ,
\label{Z}
\eeq
we find $Z \simeq 11.26$ for the ESS wave function in Fig.\ 2a.
A minimum-uncertainty solution of Eq.\ \rf{RLunc}
would have $Z=0$.

Another interesting issue is the relationship between
the ESS parameters $\al$, $\be$, $\ga_0$, $\ga_1$, $\de$
and quantities determining the shape of the corresponding
classical orbital ellipse,
namely,
the semimajor axis $a$,
the angle $\et$ it makes with the $x$ axis,
and the eccentricity $e$.
The Ehrenfest theorem requires that the mean values of the
position $\vev{\vec r}$ and the momentum
$\vev{\vec p}$ evolve according to the equations
$d\vev{\vec r}/dt = \vev{\vec p}$ and
$d\vev{\vec p}/dt = -\vev{{\vec \nabla} V(\vec r)}$.
This means that the mean position moves along the
classical trajectory if
$\vev{{\vec \nabla} V(\vec r)} = {\vec \nabla} V(\vev{\vec r})$.
However,
this condition is not satisfied for the Coulomb problem.
The connection between the ESS parameters and the shape
of the corresponding classical orbit is therefore
difficult to establish analytically
and is partly a matter of definition.

If the quantum motion did follow the classical trajectory,
knowledge of the initial position and momentum
would suffice to determine it.
One useful approximate relationship between the
ESS parameters and those of the classical orbit
in cartesian coordinates
is therefore given by taking ESS expectations of
suitable quantities as the initial data
for a classical trajectory and
thereby establishing the desired relation.
This procedure has the advantage of being analytical
and providing insight about the meaning of the
ESS parameters.

For present purposes,
we choose to express the ESS parameters as functions of
$a$, $\et$, $e$, and
the widths $\De r$ and $\de$ of the ESS,
which have no direct classical counterpart.
We find
\bea
\ga_0 &=&
\fr{I_0(2\de)}{I_1(2\de)}
\fr a {2(\De r)^2}
\left( 1 + e \cos\et -
\fr {e^2\sin^2\et}{ 1 - e\cos\et} \right)
\quad ,
\\
\al &=&
2 \ga_0^2 (\De r)^2 - 1
\quad ,
\\
\be^2 &=&
\fr{(2\al + 1)^2}{4 \ga_0 (\al+1)}
(1 - e\cos\et)
\left(\fr{I_0(2\de)}{I_1(2\de)}\right)^3
\quad ,
\\
\ga_1 &=&
- \fr{2\al + 1}{2 \be (\al+1)}
e\sin\et
\left(\fr{I_0(2\de)}{I_1(2\de)}\right)^3
\quad .
\label{rels}
\eea
Note that there are two solutions to the equation
for $\be$,
reflecting the two possible directions the trajectory
is followed.
For the examples considered above,
$\et = 0$ by assumption,
corresponding to an ESS located at the outer apsidal point.

As an example of the use of these expressions,
we can address the general issue
of whether Eq.\ \rf{RLunc}
is minimized by an ESS for any shape of classical orbit.
For definiteness,
choose $\De r$ and $\De L$ to match the example
in Fig.\ 2,
and take $\et = 0$.
Then,
the expressions \rf{rels} can be used to obtain
the quantity $Z$ of Eq.\ \rf{Z}
as a function of the semimajor axis $a$
and eccentricity $e$.
Figure 4 shows a plot of the resulting lengthy
expression for $Z$,
which provides a normalized measure of closeness
to minimum uncertainty in Eq.\ \rf{RLunc}.
The value $Z=0$ corresponds to minimum uncertainty.
Figure 4 shows that $Z > 0$ for the full range of
associated classical orbits displayed in the graph.
This demonstrates more generally that the
ESS are different from the superposition
of O(3) coherent states presented in ref.\ \cite{nau1}.

\vglue 0.6cm
{\bf\noindent V. INCORPORATION OF QUANTUM DEFECTS}
\vglue 0.4cm

Most experiments studying the properties of Rydberg wave
packets are performed using alkali-metal atoms.
These have quantum defects causing energy-level shifts
away from the hydrogenic values.
It is therefore useful
to have a relatively simple and analytical theory
for alkali-metal atoms
that has eigenstates with
the attractive properties of hydrogenic eigenfunctions
such as completeness and orthogonality,
but that generates exact quantum-defect energies.
A model satisfying these criteria,
called supersymmetry-based quantum-defect theory (SQDT),
is known to exist
\cite{sqdt}.
In this section,
we outline the planar analog of SQDT
and sketch the construction of the associated ESS.
These may be of use in modeling packets produced
in alkali-metal atoms.

The physical quantum defects $\de(n,l)$ are empirical parameters
depending in general on $n$ and $l$.
We define the planar analogues
$\de(n,\vert l \vert)$ as functions of the
absolute value $\vert l \vert$,
so that a planar Rydberg series can be established
having features in common with the standard case.
As usual,
the quantum defects for large $n$ approach asymptotic values
$\de(\vert l \vert)$ that are independent of $n$.

The planar Rydberg series has the form
\beq
E_{n^\ast} = \fr {-1} {2({n^\ast}-\half)^2}
\quad ,
\label{Esqdt}
\eeq
where $n^\ast = n - \de(\vert l \vert)$.
The quantum defects partially lift the degeneracy among
states having different values of $\vert l \vert$.
The primary objective of planar SQDT is to obtain analytical
eigenfunctions with \rf{Esqdt} as eigenenergies.
This is accomplished by defining an effective radial potential
\beq
V(r) = -\fr 1 r + \fr {\vert l^\ast \vert^2 - \vert l \vert^2}
{2 r^2}
\quad ,
\label{Veff}
\eeq
where $\vert l^\ast \vert = \vert l \vert - \de(\vert l \vert)
+ I(\vert l \vert)$ and $I(\vert l \vert)$ is an integer that
depends on $\vert l \vert$.
The radial eigenstates $R_{n^\ast l^\ast}(r)$ of this potential
have the form of $R_{nl}$ in Eq.\ \rf{radwave}
but with $n$ and $\vert l \vert$
replaced by $n^\ast$ and $\vert l^\ast \vert$,
respectively.
For certain values of the integers $I(\vert l \vert)$
and in the limit of vanishing quantum defects,
these solutions admit a supersymmetric extension.

The set of planar SQDT eigenstates
$R_{n^\ast l^\ast}(r) Y_l(\ph)$,
where $Y_l$ is an eigenstate of angular momentum,
is complete and orthonormal.
We may choose $\de(\vert l \vert)$ to match the asymptotic
quantum defects $\de(l)$,
with $l \ge 0$,
for the physical alkali-metal atom.
For example,
for lithium s and p states,
$\de(0) \simeq 0.40$ and $\de(1) \simeq 0.05$.
With these values,
the planar SQDT eigenstates reproduce the planar Rydberg
series for lithium.

An interesting issue is the ESS construction
in the context of this model.
At the classical level,
the presence of the modified $1/r^2$ term
in the effective potential \rf{Veff}
means the Runge-Lenz vector $\vec A$ is no longer conserved.
The orbit precesses at
a rate determined by the quantum defect
and is no longer a keplerian ellipse.
At the quantum level,
the commutator of the Runge-Lenz vector with the hamiltonian
is nonzero,
and the SO(4) symmetry is broken.

Despite these differences,
the RSS construction can be performed
in the presence of quantum defects
\cite{rssqdt}.
For the present case,
the oscillator variables $R$ and $P$
must be modified to
\beq
R = \fr 1 r - \fr 1 {\vert l^\ast \vert^2}
\quad , \quad\quad
P = -\fr i f \left( \partial_r + \fr 1 {2r} \right)
\quad ,
\label{qdtRP}
\eeq
where $f = \vert l^\ast \vert/\vert l \vert$.
These operators obey the uncertainty relation
\beq
\De R \De P \ge \fr 1 {2f} \vev{\fr 1 {r^2}}
\quad ,
\label{qdtuncert}
\eeq
which also depends on the quantum defect.
The wave function $\psi(r)$ minimizing this relation
has the same functional form as that in Eq.\ \rf{2drss}.

An ESS $\Psi(r,\ph)$ can be formed
as a product of an RSS $\psi(r)$ and a CSS $\ch(\ph)$.
The form of the solution is like that in Eq.\ \rf{ess}.
However,
some of the parameters must be chosen differently.
We choose $\vev{p_r} = 0$ and $\vev{L} = {\bar l}$,
and we specify $\De L$ as before.
However,
$\vev{H} = E_{{\bar n}^\ast}$ and
$\vev{r} = r_{\rm out}^\ast$ differ from the previous case.
Here,
$E_{{\bar n}^\ast}$ is the energy expectation and
$r_{\rm out}^\ast$ is the outer apsidal point for a
superposition of states with quantum-defect eigenenergies.
This means the ESS is initially located at the outer apsidal
point of a precessing ellipse.
The time evolution and revival structure
depend on the quantum defects
and can be studied by expanding $\Psi(r,\ph,t)$
in the complete set of SQDT eigenstates.

In the above,
a prescription for calculating $\vev{H}$
is needed since
the SQDT potential depends on
$\vert l \vert$ and $\vert l^\ast \vert$.
We can take advantage of completeness
to expand the initial ESS as a superposition of SQDT eigenstates:
\beq
\Psi(r,\ph) = \sum_{n,l} {\tilde c}_{nl}
R_{n^\ast l^\ast}(r) Y_l(\ph)
\quad ,
\label{qdtexpans}
\eeq
where the expansion coefficients ${\tilde c}_{nl}$,
which depend on the ESS parameters,
can be determined by inversion.
The expectation value for the hamiltonian
can then be specified as
\beq
\vev{H} = \sum_{n,l} \vert {\tilde c}_{nl} \vert^2
E_{n^\ast} = E_{{\bar n}^\ast}
\quad .
\label{fixHqdt}
\eeq

\vglue 0.6cm
{\bf\noindent VI. SUMMARY}
\vglue 0.4cm

In this paper,
we have found analytical solutions
to the planar Coulomb problem,
called elliptical squeezed states,
that minimize coordinate-momentum uncertainty relations
and that move along classical keplerian orbits.
The paper also provides an extension of the analysis
to the case where quantum defects are present,
which is of experimental importance
but theoretically difficult to treat in other approaches
because quantum defects break the O(4) symmetry.

The ESS provide an analytical tool
for studying the quantum-classical correspondence
in the Coulomb problem.
They may also be used to describe
minimum-uncertainty Rydberg wave packets
excited by short laser pulses with external fields present.
To match an ESS to a Rydberg wave packet,
we initialize the ESS at the outer apsidal
point of the orbit.
The five ESS parameters are given in terms of the
expectation values of the radial position,
the radial momentum,
the energy,
the angular momentum,
and the spread in the angular momentum.

We have studied the time evolution of the ESS for
examples with different semimajor axes and eccentricities.
The wave packets move along an elliptical trajectory
with the classical keplerian orbital period.
The squeezing of the ESS is evident as the
width of the wave packet oscillates during the motion.
The ESS maintain their form for several
orbital cycles before decoherence.

\vfill
\newpage

{\bf\noindent REFERENCES}
\vglue 0.4cm

\vfill\eject

\baselineskip=16pt
{\bf\noindent FIGURE CAPTIONS}
\vglue 0.4cm

\begin{description}

\item[{\rm Fig.\ 1:}]
Sample CSS $\vert \chi(\ph) \vert^2$
(arbitrary units)
plotted as a function of $\ph$ in radians.
All three cases shown are centered on $\ph = 0$
with $\be = \vev{L} = 30$.
The angular-coordinate widths of the CSS depend on $\de$
and have the values
$\De L = 0.5$ for $\de \simeq 0.804$ (solid line),
$\De L = 1.5$ for $\de \simeq 4.757$ (dotted line),
and $\De L = 2.5$ for $\de \simeq 12.753$ (thick solid line).

\item[{\rm Fig.\ 2:}]
The modulus of an ESS $r \vert \Psi(r,\ph,t) \vert^2$
(arbitrary units)
plotted as a function of $r$ and $\ph$ at the times:
(a) $t=0$,
(b) $t = \fr 1 3 T_{\rm cl}$,
(c) $t = \fr 1 2 T_{\rm cl}$,
(d) $t = \fr 2 3 T_{\rm cl}$,
(e) $t = T_{\rm cl}$.

\item[{\rm Fig.\ 3:}]
Comparison of the modulus squared (arbitrary units)
of two ESS with different
eccentricities at the time $t = \fr 1 2 T_{\rm cl}$.
Both wave packets have ${\bar n}=45$ and $\De L = 2.5$.
and are viewed from a point
on the positive $x$ axis looking in toward the origin.
For Figure (a), ${\bar l}=30$,
while for Figure (b), ${\bar l}=40$.

\item[{\rm Fig.\ 4:}]
A plot of the quantity
$Z = (\De A_x \, \De A_y
- \vert \vev{HL} \vert )/\vert \vev{HL} \vert$
as a function of the semimajor axis $a$ in atomic units
and the eccentricity $e$ of the associated classical orbits.

\end{description}

\vfill\eject
\end{document}